# Direct strain and elastic energy evaluation in rolled-up semiconductor tubes by x-ray micro-diffraction

A. Malachias[1,2*], Ch. Deneke[3], B. Krause[4], C. Mocuta[2], S. Kiravittaya[1], T. H. Metzger[2], O. G. Schmidt[1,3]

[1]*Max-Planck-Institut für Festkörperforschung, Heisenbergstrasse 1, D-70569 Stuttgart, Germany*

[2]*European Synchrotron Radiation Facility, BP 220, F-38043 Grenoble Cedex, France*

[3]*Institute for Integrative Nanosciences, IFW Dresden, Helmholtzstr. 20, D-01069 Dresden, Germany*

[4]*Institut für Synchrotronstrahlung, Forschungszentrum Karlsruhe, Postfach 3640, 76021 Karlsruhe, Germany*

## Abstract

We depict the use of x-ray diffraction as a tool to directly probe the strain status in rolled-up semiconductor tubes. By employing continuum elasticity theory and a simple model we are able to simulate quantitatively the strain relaxation in perfect crystalline III-V semiconductor bi- and multi-layers as well as in rolled-up layers with dislocations. The reduction in the local elastic energy is evaluated for each case. Limitations of the technique and theoretical model are discussed in detail.

* e-mail address: amalachias@lnls.br

PACS: **61.05.cp, 61.46.-w**

Present address: Laboratório Nacional de Luz Síncrotron

C.P. 6192 – Campinas, Brazil.

## I. Introduction

The ability to release and transfer semiconductor layers with high crystalline quality has lead to novel possibilities for the fabrication of devices on the micro- and nano-scale [1]. Strain properties can be exploited to produce bending of the layer, leading to a repositioning of a predefined film area [2, 3] or curling into rolled-up tubes [4, 5]. In these structures the partial release of the flat layer strain results in a significant change of the lattice configuration, modifying properties such as semiconductor band gap energies [6, 7] and charge carrier mobilities [8, 9]. Rolled-up semiconductor micro-/nanotubes can be used as flexible ring resonators [10-12] as well as on-chip integrative refractometers [13], linear fluidic devices [14, 15] and mechanical components [16, 17].

A general layer design that is often used for producing rolled-up structures consists in the heteroepitaxy of two or more pseudomorphically grown thin films on top of an etchant sensitive layer. Selective etching is then used to release the top layers that relax elastically by rolling up into a tube as depicted in Fig. 1(a). Since the process depends only on the preexistence of a strain gradient across the layers [18] it can be generalized for obtaining a heterostructure that combines different compounds such as organic/semiconductor [19], oxide/semiconductor [12], metal/semiconductor [20, 21], or combinations thereof [2] into a self-assembled radial multilayer. Such possibility is, in fact, one of the driving forces for research in this field. However, for applications in material integration and optoelectronic structures the precise knowledge of local strain in rolled-up layers is crucial for band gap engineering, as well as for the fine tuning of strain dependent electric and magnetic properties and to layer-to-layer interface optimization.

Structural characterization of single rolled-up tubes has been carried out mainly by microscopy methods such as scanning electron microscopy (SEM) and transmission electron microscopy (TEM). SEM is generally the technique of choice to provide insights on the tube radius, morphology and layer folding quality [2] while TEM has been successfully used to study interfaces between successive windings [19-21]. Average strain can be obtained from micro-Raman measurements, without a clear distinction of the strain components in each direction [22-24]. Despite of the information available from these techniques a complete scenario describing the strain relaxation inside tubes and correlating its mesoscopic and crystalline properties could only be drawn recently by using x-ray microdiffraction, that allows for directly measuring the radial lattice parameter distribution over the rolled-up layers [25].

For this work $In_xAl_yGa_{1-x-y}As/GaAs$ ($x\leq0.33$, $y\leq0.2$) epitaxial layers were grown on top of an etchant sensitive (sacrificial) AlAs layer on GaAs(001) substrates [26]. After rolling, the lattice parameter configuration inside single tubes is retrieved by x-ray micro-diffraction. The paper is organized as follows. Firstly we discuss a model based in continuum elastic theory which is used to obtain the local strain status in a rolled-up tube. The result is compared with models available in the literature. Secondly we describe the sample design and use of x-ray micro-beam as a probe to study strain relaxation in single tubes. The reciprocal space configuration and alignment procedure are discussed. A simple x-ray model that holds for extracting quantitative information is shown. Finally x-ray measurements performed in rolled-up III-V semiconductor tubes are simulated, allowing the direct evaluation of local strain and elastic energy.

## II. Elasticity theory

The rolling-up of pseudomorphically strained thin films is driven by the minimization of the total elastic energy $E$ as described in refs. [25-29]. For epitaxially grown cubic crystals without torsion components $E$ is locally given by [27, 28]

$$E(r) = \frac{C_{11}(r)}{2}(\varepsilon_x^2 + \varepsilon_y^2 + \varepsilon_z^2) + C_{12}(r)(\varepsilon_x\varepsilon_y + \varepsilon_x\varepsilon_z + \varepsilon_z\varepsilon_y), \qquad (1)$$

where the subscripts $x$, $y$ and $z$ indicate directions that are parallel to the main crystallographic axes. The dependence of the elastic constants on the local composition at the position $r$ is explicitly shown in eq. 1 for $C_{11}$ and $C_{12}$ and implicit for the strain components. In the following paragraphs the subscripts $x$, $y$ and $z$ for the planar layer will be referred, respectively, as $t$ (tangential), $l$ (longitudinal) and $r$ (radial) for clarity.

If such an epitaxial single crystalline film is curved in a defect-free cylinder with inner radius $R_i$ the lattice parameter in the tangential ($a_t$) direction varies continuously inside the layers. The local value of $a_t$ at the position $r$ with respect to the tube center is given by [29]

$$a_t(r) = a_i\,(1 + r/R_i), \qquad (2)$$

where $a_i$ is the tangential lattice parameter of the inner surface as shown in Fig 1(b). After the releasing of the layers the crystalline lattice is allowed to expand (or contract) in the radial ($z$) direction and the three strain components are then related by the plane strain condition

$$\varepsilon_r = -[C_{12}/C_{11}](\varepsilon_t + \varepsilon_l) \quad (3)$$

It is possible, hence, to obtain the final configuration for a given rolled-up tube by minimizing the total elastic energy for its layer structure. By replacing equations (2) and (3) into (1) one can re-write the total elastic energy for a stack of $N$ layers as

$$E_{tot}(a_i, R_i, \varepsilon_l) = \sum_{n=0}^{N-1} \int_{d_n}^{d_{n+1}} E(a_i, R_i, \varepsilon_l, r)dr \quad , \quad (4)$$

where the index $n = 0$ denotes the bottom interface of the first layer in the stack ($d_0 = 0$) and $d_n$ are the positions of the interfaces in the layer stack using $d_0$ as a reference. For a fixed longitudinal strain $\varepsilon_l$ the values of $R$ and $a_i$ that minimize $E_{tot}$ for an atomic chain aligned perpendicularly to the tube surface, as sketched by the dashed red box in Fig. 1(a), can be found numerically by evaluating eq. 4 in a range of the configuration space of $a_i$ and $R_i$. Figure 1(c) shows a 3D plot of $E_{tot}$ for a selected bilayer configuration consisting of 200Å $In_{0.2}Ga_{0.8}As$ / 200Å GaAs (therefore $N = 2$, $d_1$= 200Å, $d_2$= 400Å) with $a_l = a_{GaAs}$ =5.653Å in a limited configuration space window. The position of the minimum indicates the equilibrium state for the tube, which is obtained by the conditions

$$\partial E_{tot}/\partial a_i = 0 \quad \text{and} \quad \partial E_{tot}/\partial R_i = 0 \quad . \quad (5)$$

Analytical solutions for N = 2 and multilayer cases are given in references 27 and 29. The values of $a_i$ and $R_i$ obtained here from direct energy minimization using eq. 4 via numerical methods are in quantitative agreement with those obtained from the analytical solution from reference 22 within an error bar of ±1%. Nevertheless, a numerical minimization of eq. 4 allows additionally for an evaluation of the equilibrium conditions with any combination of fixed parameters (e.g. fixing $R_i$ to the experimentally observed radius), instead of restricting the process to $a_i$ and $R_i$ from the predicted equilibrium.

## III. Experimental

### *A. Samples and layer layout*

Selected samples were chosen to evidence how different layer configurations affect the lattice relaxation inside rolled-up tubes. All samples were grown on top of 200Å AlAs etchant-sensitive layers deposited on GaAs(001) substrates. For the first sample – referred here as '*bilayer*' – 185Å of $In_{0.2}Al_{0.2}Ga_{0.6}As$ and 185Å of GaAs were deposited on top of the AlAs layer. In this bilayer the In mainly determines the strain while the replacement of Ga atoms by Al atoms slightly modifies the elastic constants. In the second sample, four layers were stacked on top of the AlAs film by repeating twice the deposition of a bilayer structure of 200Å of $In_{0.2}Ga_{0.8}As$ and 300Å of GaAs. This sample, referred here as '*quad-layer*', was designed to probe the possibility of non-monotonic radial strain relaxation across the interfaces. Finally for the last sample, that we named '*dislocated*', a bilayer structure of 250Å of $In_{0.33}Ga_{0.67}As$ and 250Å of GaAs was grown. Since the critical thickness for the ternary alloy film with 33% of In atoms in the III site is smaller than 40Å [30] a large density of defects is expected and, therefore, a different relaxation after rolling. The layer layouts for the bi-, quad-layer and dislocated layers are schematically represented in the left panels of Figs. 2 (a-c), respectively.

The right panels of Figs. 2(a), 2(b) and 2(c) show x-ray reflectivity measurements performed on the corresponding reference flat layers, that are sketched in the left panels, as a function of $q_r = (4\pi/\lambda)(\sin(2\theta/2))$, where $\lambda$ is the x-ray wavelength and $2\theta$ the detector angle. The nominal grown layer thicknesses are compared with the values of x-ray reflectivity simulations [31] for the flat layers and (004) x-ray diffraction from the rolled-up layers in table I. For the flat layers the differences between nominal and measured thickness are inside the error bars from the growth and fitting processes. The small discrepancies obtained in the rolled-up layers will be addressed furtherer in the text for each case.

For the bilayer sample the rolled-up tubes were lithographically positioned by a two step etching procedure. As shown in the left panel of Fig. 3(a) 200μm stripes followed by 300μm spacers were defined along the [100] direction. The topmost GaAs layer was then removed in the spacers by $H_3PO_4:H_2O_2:H_2O$ (1:10:500) shallow wet etching [7]. In a second photolithography step the underneath AlAs layer was laterally exposed by deep narrow trenches along the [010] direction obtained after etching in a HBr(50%) :$K_2Cr_2O_7$(0.5mole/L): $CH_3COOH$(100%) (2:1:1) solution [15, 32]. Finally the layers were released by etching the AlAs layer with diluted HF(50%):$H_2O$ (1:10) solution for 40s. The resulting 200μm-long tubes have an inner radius of 1.3±0.1μm and performed about 10 rotations as shown by the SEM image of the inset of Fig. 2(a). In the shallow etched areas the film does not perform rotations and only produces wrinkles on the surface [33]. The preparation procedure employed for the bilayer was optimized for producing tubes with up to 10 rotations, minimizing the occurrence of cracks along the tube, in order to explore the effect of multiple rotations in the strain profile of the layers.

In the quad-layer sample, long (500μm) deep trenches were defined lithographically to minimize tube cracks and tubes with 14±0.3μm radius and a maximum of 2 windings were obtained.

Finally for the dislocated sample the trenches were obtained from surface scratching and long rolled-up tubes were produced exhibiting a radius of 1.5±0.1μm with 5~6 rotations. For all samples the tubes roll along one of the <010> directions. Tube openings are shown in the SEM images of the insets for all cases.

*B. X-ray diffraction from single tubes*

The x-ray microbeam experiment was performed at the ID01 beamline of the European Synchrotron Research Facility (ESRF) by using Be compound refractive lenses (CRL). The focused x-ray spot achieved at 8.8KeV ($\lambda$ = 1.409Å) has a size of 6×6μm at the sample position and a divergence of 0.05º. Such spot size is small enough to measure diffraction from single tubes. The flux density gain after the CRL is of approximately 5000 times, allowing for measurements in very small sample volumes. Finally, the divergence is obtained by measuring the Si(004) peak width of an analyzer Si(001) crystal. Diffraction measurements are performed using an avalanche photodiode as detector. An optical microscope aiming at the center of the a 4+2 circle diffractometer allows for a view of the sample surface and optical pre-alignment in which the longitudinal axes of the tubes are oriented perpendicularly to the x-ray beam path as represented in Fig. 3(a).

Diffraction measurements are then performed around the GaAs (004) reciprocal space position for all samples. The fine x-ray beam positioning on the sample can be easily performed by taking profit of the tube geometry. Since the crystalline layers inside the tubes have a radial symmetry it is possible to suppress the diffraction from the substrate by de-tunning the substrate lattice from the specular θ-2θ condition as represented in Fig. 3(b) [25].

A sketch of the reciprocal space diffraction intensity distribution for the rolled-up and flat layers is shown in Fig. 3(c). While the diffraction of the flat layers consists in a very localized spot in the reciprocal space the diffraction from the curved crystals can be observed along a powder-like rim of intensity. Hence, to enhance the sensibility to tube diffraction a detuning is performed in the sample angle θ by adding an increment Δ of about 15º as shown in Fig. 3(b). Although the procedure can be also performed with a negative detunning the incoming flux that illuminates the tube spreads, creating a larger footprint and therefore producing a less intense diffraction signal despite of the reduced background.

The tubes can then be found by laterally scanning the sample with the detector 2θ angle fixed at the GaAs or $In_xAl_yGa_{1-x-y}As$ (004) rolled-up layer reciprocal space position [34]. For such condition the angles define a point in reciprocal space that is sensitive to diffraction of rolled-up material solely. Consequently, a strong signal is observed when a tube is on the beam, in contrast with the absence of counts obtained from the flat film regions, shallow etched and wrinkled areas. An inspection of both panels of Fig. 3(a) shows the correspondence between an optical image and the scanning x-ray diffraction performed in the same area of the bilayer sample.

Once a tube of interest is selected some optimization on the alignment is required. A preliminary radial scan is performed by spanning solely the detector (2θ) angle, which allow for the observation of diffraction peaks from the layers inside the tube. The angle 2θ is then fixed to one of the peaks and the translation stages can be scanned to optimize the diffracted intensity (i.e. bring the tube into the center of the x-ray spot). Finally, it is crucial to check whether the beam is perpendicular to the tube axis by performing an azimuthal scan (hereafter referred as ϕ-scan). As it will be shown in the next session $\phi$-scans reveal an interplay between layer size, tilting of the layers inside the tube and, therefore, the packing quality of successive windings.

## IV. Results and Discussion

### *A. Measurement and azimuthal alignment*

Radial scans performed in a rolled-up tube of the bilayer sample and on the reference flat layers are shown in Fig. 4. Both scans were performed in the vicinity of the GaAs (004) reflection and their corresponding paths in reciprocal space are represented in Fig. 3(c). The inset of Fig. 4 shows a schematic representation of the sections in a rolled-up tube that contribute to diffraction at the (004) position. The main contribution comes from the two opposite sectors in which the radial lattice planes are oriented perpendicularly to the momentum transfer vector [25]. Hence, in a detuned radial scan the radial lattice parameter profile is measured along the ⟨004⟩ direction. It is possible to draw preliminary qualitative conclusions by simple inspection of both measurements. In the flat system the GaAs layer is completely unstrained while the $In_{0.2}Al_{0.2}Ga_{0.6}As$ layer is under a biaxial compressive strain $\varepsilon_{//}$ imposed by the host GaAs(001) substrate. An out-of-plane expansion given by $\varepsilon_z=-2(C_{12}/C_{11})\varepsilon_{//}$ is then observed for this In-rich layer ($\varepsilon_{//} = \varepsilon_x = \varepsilon_y$ for the flat film), leading to an out-of-plane lattice parameter of 5.822Å. Curving the layers into a tube will allow for a partial in-plane relaxation of the $In_{0.2}Al_{0.2}Ga_{0.6}As$ film along the tangential direction as depicted in Fig. 1(b), leading to an out-of-plane – radial for the curved layers – contraction for a fixed $\varepsilon_l$. As a result, the diffraction peak for the rolled-up layer shifts to larger $q_r$ values with respect to the corresponding position for the flat layer. On the other hand, since a tangential expansion is expected along the tube wall for the given $a_l = a_{GaAs}$, the out-of-plane GaAs layer lattice also contracts slightly. Therefore, the corresponding lattice parameter difference between the two layers decreases, which is observed as an approaching of the two peaks in Fig. 4.

Prior to a detailed interpretation of the measurements and introduction of an x-ray model some comments should be made concerning the reciprocal space profiles observed by performing an azimuthal tube alignment. Considering the beam divergence of 0.05º the tangential section of a tube where the lattice is aligned perpendicularly to the momentum transfer vector has a very reduced dimension of about 10Å. Since the beam spreads laterally along its 6μm size the useful footprint for diffraction is of about $0.006\mu m^2$ as represented in Fig. 5(a). Such dimensions are consistent with relative diffraction intensities calculated with respect to the incoming flux and the diffraction from flat layers and make the use of a focused beam mandatory for recording reasonable signals. For the footprint geometry shown in Fig. 5(a) the $\phi$-dependent diffraction width is related to three factors: (i) the tube radius $R_i$, which implies that in small diameter tubes with pronounced curvatures the diffracted intensity should be more sensitive to the $\phi$-alignment (narrower profile); (ii) the folding faults and tilting of neighbor windings, that would produce a mosaic spread of scattering due to imperfections in the matching of internal walls. A larger number of these faults would then be proportional to the number of windings *w*; and; (iii) the size broadening due to the finite layer thickness.

Figure 5(b) shows $\phi$-scans obtained at the fixed reciprocal space position for the tube GaAs peak of Fig. 3 in tubes with different total layer thickness. For GaAs layers with 185Å and 300Å thickness (the last not further explored here) embedded in bilayer tubes with a large number of rotations – 10 and 7, respectively – and inner radius of 1.5μm and 2μm, are obtained, respectively. The peak of the 300Å GaAs layers of a quad-layer tube – with an inner radius of 7μm and only one layer turn – is also shown. Although the exact interplay between folding and tilting mosaic and the tube radius cannot be directly evaluated from the widths of these curves one can infer an upper bound of the effective overall mosaic spread for the layers, which includes the previous factors, by deconvoluting the angular width of these curves with their expected sizes broadenings. In such approach the effect of the radius $R_i$ is assumed as considerably smaller than the mosaic/tilt induced broadening.

For the given diffraction geometry the calculated thickness-dependent angular width $\Delta\phi$ of a 185Å thick layer is of 0.235°, while for a 300 Å thick layer it would correspond to $\Delta\phi = 0.157°$. The overall layer mosaic spread M for the three tubes shown in Fig. 5(b) can then be calculated as $M = (\delta^2 - \Delta\phi^2)^{½}$ [35], where $\delta$ is the measured angular width of a $\phi$-scan. For the larger profile of the 185Å layer M = 0.99° was obtained. This value is slightly smaller compared to the 300Å bilayer tube, with M = 0.62°, and much smaller for the 300Å quad-layer tube, that exhibits M = 0.29°. In all cases discussed above the beam divergence is much smaller than the obtained mosaic spread values.

A comparison of M values for these three tubes indicates that there is a much stronger dependence on the overall layer mosaic spread on folding and tilting faults [factor (ii) described above] than on the tube radius. A multirotation tube formed from a thinner layer is susceptible to develop more layer tilts or loosely packed windings with respect to a thicker layer due to its fragility. Such behavior is shown by the narrow $\phi$-scan profile measured in the 1000Å wall thickness of the quad-layer tube. Despite of having a much larger radius (14μm), which would induce a broader profile, the layer stack is quite robust against tilting and folding faults. Scans in $\phi$ performed at the $In_xAl_yGa_{(1-x-y)}As$ peak have shown identical profiles. Hence, a $\phi$ alignment is always needed to optimize the diffracted intensity by tuning the preferred packing orientation of the tube windings and reveals quantitative information on the average mosaic spread of the layers.

*B. X-ray model and fit precision*

In order to quantitatively analyze the diffraction from the tubes produced from the layer systems of Fig. 2 one must introduce a convenient x-ray model. Such model will be based on three assumptions:

(i) The reduced layer thickness allows for the use of kinematical theory;

(ii) the diffracted intensity is mainly sensitive to the scattering from a region of the tube where the radial lattice parameter is aligned to the momentum transfer vector, and, therefore, probes mainly the $a_r$ lattice profile;

(iii) the total diffraction intensity measured is an incoherent sum of the intensities of all W turns due to the random crystal misalignment between successive windings and the formation of a thin oxide layer at the interfaces [19];

(iv) the tube is homogeneous along its longitudinal direction on a length scale of the order of the x-ray beam size (6μm) which is used to probe its properties.

The q-dependent diffraction intensity observed from a multirotation tube will be then given by

$$I(q) = I_0 \sum_{w=0}^{W-1} \left| \sum_{n=0}^{N-1} \left( \sum_{j=0}^{A_n} f_n e^{iqr_j} e^{-q\frac{\sigma^2}{2}} \right) \right|^2 , \qquad (6)$$

where the summation over *w* accounts for the incoherent diffraction of the successive windings, the summation over *n* accounts for the layer stack in each wall and the summation over *j* for the atom positions in each layer $r_j$ along the tube radial direction [36]. In this equation $f_n$ and $A_n$ are the average effective atomic scattering factor and the number of atoms in the layer *n*, respectively; σ represents an overall layer roughness. The input parameters for a calculation of the diffraction curve using eq. 6 are σ, $f_n$ and the atomic positions. These later are obtained by the following procedure. Firstly the constants $R_i$ and $a_i$ are extracted by minimizing the elastic energy for the first turn. Then, using eq. 2 the tangential lattice profile is obtained for all positions inside this layer stack. The radial lattice parameter ($a_r$) profile as a function of the position in the tube is then generated by applying eq. 3. Similar $a_r$ profiles are also obtained for $R_w = R_i + wD$, where D is the total layer stack thickness ($D = \Sigma\Delta d_n$). Finally, the calculated atomic positions are used as input for eq. 6 and a simulation of the diffraction profile is obtained.

Figure 6(a) shows the best fit found for the data recorded on the bilayer sample. This fit was obtained for an $In_{0.2}Al_{0.2}Ga_{0.6}As$ (150Å) / GaAs (190Å) bilayer with $R$ = 1.30μm, σ = 20Å, $a_l = a_{GaAs}$ (no longitudinal relaxation) and the nominal In and Al concentrations. Identical diffraction profiles were found in different tubes rolled-up from the same sample. The reduced layer thickness obtained in the diffraction simulation can be attributed to a thin oxide formation after the layer release [19]. Particularly for the $In_{0.2}Al_{0.2}Ga_{0.6}As$ layer, a more pronounced difference with respect to the nominal and measured flat layer thickness shown in Table I is obtained. Nevertheless, assuming the formation of a native oxide with maximum thickness of 20Å for the Al-rich layer, the deviations from the thickness obtained from different methods lie inside the estimated error bars. Corresponding lattice parameter profiles for the inner, middle and outter windings are shown in Fig. 6(b). The fixed longitudinal lattice parameter [dashed line in Fig. 6(b)] indicates that no relaxation takes place in this direction during rolling or after it, evidencing that a strong, wafer-like layer bond takes place between adjacent windings.

The fitting values used for the bilayer tube indicate a very good agreement with the continuum elastic theory as also shown in other tubes explored in ref. [25]. It is worth to probe the actual error bars on the fitting procedure and its dependence on each of the modeling parameters. This is shown schematically in Fig. 6(c). Although equations 2-5 interconnect the values of the fitting variables $R_i$, $a_i$, ε's and a's we depict the effect of changing independently one parameter in the fit while keeping the others fixed in their optimized value. Such procedure establishes, semi-quantitatively, intrinsic error bars of the x-ray diffraction modeling. For the bilayer the most relevant parameters are: (i) $In_{0.2}Al_{0.2}Ga_{0.6}As$/GaAs layer thickness; (ii) In concentration ($C_{In}$) – here we assume that the Al influence on strain is negligible for the concentration used; (iii) longitudinal lattice parameter $a_l$ (or alternatively the longitudinal strain $\varepsilon_l$) and; (iv) tube inner radius $R_i$.

Maybe the most evident parameter is the layer thickness ratio of item (i). An incorrect balance between $D_{GaAs}$ and $D_{InAlGaAs}$ renders the intensity of one peak larger than the other, as shown by the red curve of Fig. 6(c). Additionally, an error in the total thickness changes the strain distribution and, therefore the position of the peaks in $q$, that drift in opposite directions. From the fitting procedure it is possible to estimate the error bar for this parameter as small as 15Å.

A change in the In concentration – item (ii) – will displace both peak positions laterally in opposite $q$ directions as shown by the green curve due to a change in strain. Figure 6(d) depicts the changes observed in the lattice parameter profile by increasing the In concentration $C_{In}$ by 1%. The effective error bar for $C_{In}$ was found to be of about 0.5%. The Al content ($C_{Al}$) was also varied in our simulations (not shown). Although it introduces no strain the presence of Al atoms in the layer alters its elastic properties. $C_{Al}$ has proven to effectively change the fit quality only for a concentration that differs from the nominal by more than 10%.

Changing $a_l$ (or $\varepsilon_l$) – item (iii) will move both peaks laterally on the $q$ axis in the same direction. In the case $a_l > a_{GaAs}$ the longitudinal expansion of the lattice parameter will produce a contraction in $a_r$ for the GaAs layer. For the In-rich layer a contraction in $a_r$ is also observed and therefore both peaks are shifted to higher values of $q$. An effective error bar of $\Delta\varepsilon_l \cong 0.08$ was found. Changes on this condition will be further explored while fitting the quad-layer and dislocated tubes (sections IV.C and IV.D, respectively).

Finally it is worth to analyze the effect of a change in $R_i$. As shown by the blue curve of Fig. 6(c) and the lattice parameter profile in Fig. 6(e) a 10% larger value of $R_i$ will split the peaks apart while smaller $R_i$ values will bring them closer. For the given bilayer tube the fit is sensitive to variations in $R_i$ of about 4%. Here one must notice that, as shown in Fig. 5(a), the x-ray diffraction method probes a very reduced volume of the rolled-up crystalline layers. The value of $R_i$ obtained by this method represents the local curvature of the illuminated tube area. However, by measuring several tubes and/or several positions along one tube it is possible to average the diffraction profile and extract a realistic $R_i$. In round and well-packed tubes like the ones used in this work radial scans performed in different positions or different tubes are very similar. Finally, the values of $R_i$ obtained by all fits are in very good agreement with direct methods like SEM and TEM, showing that local curvature and radius coincide for our structures.

A quick exploration on the effect of the incoherent sum given by eq. 6 is also performed. The lower solid black line in Fig. 6(c) corresponds to the expected diffraction profile for a completely coherent lattice matching in each 2 turns in the bilayer tube. For such system the summation in *w* (windings) of eq. 6 is set inside the squared modulus. The result is a curve with discrete peaks separated by the reciprocal of the bilayer thickness ($2\pi$/D) with an external envelope given by the fully incoherent fit of Fig. 6(a). A more realistic situation where the lattice of only two of the windings are coherently matched and all other turns interfere incoherently is represented by the orange solid line of Fig. 6(c). Although a completely crystalline bonding through the whole rotation is unexpected this situation would represent a tube where a considerable amount of large aligned crystalline bonded areas can be found. Considering the effects of roughness and beam divergence the interference minima of the wiggling would be less pronounced, although still measurable. No evidence of coherent layer matching in large areas between neighboring windings was found for any of the tubes that were measured, as well as for the tubes from ref. [25]. This can be explained by the imperfect crystal lattice matching in the interfaces of windings as well as by the formation of a thin amorphous oxide layer [19].

### *C. Multilayer tube*

The degree of partial strain relaxation that takes place after the rolling process is strongly dependent on the layer stack. While for bilayer tubes the resulting strain profile after roll indicates a good agreement with the continuum elasticity description employed in the last session it is worth to understand in which extension it may be also applied in a system where a non-monotonic relaxation may take place. Figure 7(a) shows a radial scan performed in the flat reference layers for the quad-layer sample. The larger number of minima and maxima with respect to the bilayer flat films of Fig. 4 is generated by the double repetition of the 300Å GaAs / 200Å $In_{0.2}Ga_{0.8}As$ stack. As observed for the bilayer sample the peak of the In-rich layer corresponds to an out-of-plane lattice parameter of 5.82Å ($q_r$=4.318Å$^{-1}$), expected for a 1.4% in-plane strained film with the nominal 0.2 In content.

A radial scan on the rolled-up tube is shown in Fig. 7(b). Despite of the curvature of the surface the measured profile (open dots) exhibit several deep sharp minima as for the flat layers, indicating that the layers scatter as very softly bowed structures due to the small extension of the beam tangential footprint. A first attempt to simulate the observed x-ray profile is shown by the red dashed line. Assuming that no relaxation takes place along the longitudinal direction of the tube one cannot reproduce the exact peak position and/or their relative intensities. Since the tube diameter, calculated [22] and also observed as 14μm, is much larger than in the bilayer case the wafer bonding between successive windings will only occur after a long extension of flat material has been already relieved from the substrate constraints. The much larger perimeter of the cylinder implies a reduced influence of the un-etched film in the final tube strain status. It is, therefore, reasonable to assume that some longitudinal relaxation occurs. In fact, the best fit of the diffraction profile can only be obtained by fixing a longitudinal lattice parameter intermediate between the GaAs and $In_{0.2}Ga_{0.8}As$ bulk values [shown in Fig. 7(c)]. Such value corresponds to a longitudinal strain of 0.7%, compressive for the In-rich layers and tensile for the GaAs layers. By assuming the longitudinal strain above the external envelope of the curves shifts into the larger $q_r$ direction, and a more suitable balance between the sharp peak intensities is found. This shift of the fit curve cannot be obtained by changing any other parameter in the simulation. An effective error bar for the longitudinal strain determination can be estimated as ±0.2% for this case.

The layer thickness used in the simulations of Fig. 7(b) for the GaAs inner and outter layer are 280Å and 270Å, respectively. For the $In_{0.2}Ga_{0.8}As$ layers we have obtained 190Å (inner layer) and 180Å (outter layer). The tube radius found by the minimization of the elastic energy given by Eq. 4 was $R_i$ = 6.94μm, in agreement with the value found using the evaluation method of reference [22] (6.88μm). An average roughness $\sigma$ = 30Å was used for the fitting.

Figure 7(c) shows the lattice parameter profile obtained from the fit of Fig. 7(b). By a simple inspection of the less pronounced slopes of tangential and radial lattice parameter profiles it is possible to infer that the releasing of the elastic energy in this tube by rolling is much less effective than in the bilayer tube. However, the longitudinal lattice parameter relaxation adds another degree of freedom for the energy minimization, leading to a structure where the final local stored elastic energy is similar to the bilayer case. A more quantitative and detailed discussion will be done in Section E.

*D. Dislocated layer tube and effective strain*

A limiting case for the use of the continuum elastic model shown would be a tube in which one or more layers have a large density of defects. For such tubes the possibility of rolling as a strain releasing process is still possible [37]. The tube rolling can be used, then, to probe the effective strains which are stored in these layers [38]. The dislocated layer rolled here is obtained from a $In_{0.33}Ga_{0.67}As$/GaAs bilayer in which the In-rich film thickness (~250Å) is much larger than the critical thickness for this given concentration [~40Å, see ref. 30]. A density of defects of about $10^6$~$10^7$cm$^{-2}$ is expected [39].

Figure 8(a) shows the measured diffraction intensity in the vicinity of the GaAs (004) reciprocal space position for the flat and rolled-up dislocated tube layers. A quick comparison of the radial scan in the dislocated flat film and the scan performed in the defect-free bilayer sample of Fig. 4 evidences a considerably larger roughness of the first [37]. Despite of the large number of dislocations the $In_{0.33}Ga_{0.67}As$ peak for the flat layer ($a_\perp$ = 5.899Å) corresponds to a pseudomorphically strained InGaAs alloy with 30% In content, indicating that only a reduced fraction of the 2.2% in-plane strain is released by defect formation. A radial scan performed in grazing-incidence geometry directly shows an in-plane average lattice parameter of 5.695Å for the dislocated bilayer. Such value corresponds to a relaxation of 0.7% with respect to a hypothetical coherent biaxially strained $In_{0.33}Ga_{0.67}As$ film on GaAs.

Two fits of the dislocated tube diffraction are shown in Fig. 8(b). Particularly in this case, in which plastic deformations take place, the use of nominal parameters of layer thickness and strain, assuming a defect-free tube, lead to an extremely different lattice parameter distribution inside the rolled-up layers and is unable to model the observed diffraction profile. By employing the energy minimization method described in section II, as well as the solutions proposed by references [22, 28] one obtains a tube radius of 1μm, much smaller than the observed 1.5μm. Alternatively, for the layer configuration of 250Å $In_{0.33}Ga_{0.67}As$ / 250Å GaAs a tube with the observed 1.5μm radius can only be obtained from the methods of section II by assuming an In content of 0.22 and the corresponding strain of 1.5% given by such concentration. Such value corresponds directly to the 0.7% relaxation obtained in the in-plane (400) diffraction. Therefore, the observed radius of a tube with dislocated layers provides already a fairly good estimation of the amount of strain, which is released by the formation of defects.

By fixing the radius R = 1.5μm in eq. 4 a reasonable lattice parameter configuration can be found. However, although the use of the observed radius accounts in part for the strain released during layer deposition an In content of 0.33 must be used to generate the correct lattice parameter profile which fits the tube diffraction. This finding can be explained by the dimensions of the diffracted beam footprint, discussed in section IV-A. For the dislocation densities expected here only one or two defects per bilayer turn will be illuminated in the 0.006$\mu m^2$ effective diffraction area. Therefore, the local effects of the strain induced due to the In-rich alloy will prevail over the presence of dislocations. In other words, such result also suggests that the volume of material in the layers affected by the presence of dislocations is about two orders of magnitude smaller than the volume which remains unaffected by the defect strain field and is probed by the x-ray beam. The orange solid line of Fig. 8(b) is then obtained by minimizing eq. 4 for a 230Å $In_{0.33}Ga_{0.67}As$ / 225Å GaAs bilayer, with fixed $R_i$ = 1.5μm, $C_{In}$ = 0.33 and $\sigma$ = 45Å.

Finally, a longitudinal lattice parameter of $a_l$ = 5.7Å must be used in eq. 4 in order to bring the simulated peaks into the correct $q_r$ positions as represented by the blue solid line of Fig. 8(b). The use of this value corroborates the initial strain status of the layers and indicates, as for the bilayer sample of section IV-B, that an inter-layer wafer-like bonding in small radius tubes keeps the longitudinal strain status of the flat films. The final lattice parameter profile obtained is shown in the inset of Fig. 8(b).

*E. Local elastic energy and partial relaxation*

Although the relaxation of the longitudinal lattice of two of the tubes presented here accounts for a small change in the radial lattice parameter profile and, therefore, the observed tube diffraction, it strongly influences the final elastic energy stored inside the rolled-up layers. It is worth, therefore, to understand how the elastic energy varies locally inside the tube walls.

Applying eq. 1 to the lattice parameter profile of the bilayer tube shown in Fig. 6(b) leads to the elastic energy profile of Fig. 9(a). The energy stored in the flat pseudomorphically strained $In_{0.2}Al_{0.2}Ga_{0.6}As$ film, represented by the black horizontal line, is partially released during roll. In the flat layers the In-rich film stores a per atom energy of 3.28meV, while the total energy in the bilayer, evaluated for an atomic chain perpendicular to the surface is of 353meV. After the rolling of the layers part of the initial energy is redistributed between the GaAs and the In-rich layers, while another part is released in the rolling process. Although after the formation of the tube most of the energy (~85%) remains in the In-rich film due to the absence of longitudinal relaxation in this tube a small fraction of about 15% of the total final energy is stored in the GaAs layer. The energy difference between the inner and outter layer, also shown in Fig. 9(a) is very small. While the radially integrated energy for an atomic chain of the inner winding is equal to 160meV it reaches 164meV after 10 turns. Despite of such difference of about 2.5% the positions of the energy minima of each layer shift outwards as the number of turns increase. The comparison of integrated elastic energies along atomic chains discussed here for the flat and rolled-up layers provides an estimation of the magnitude of the elastic driving forces involved on the rolling process. This is represented by the energy difference between the flat and rolled-up cases.

In the quad-layer a more complex energy redistribution is found. The initial energy of 3.48meV/atom stored in the In-rich flat layers will be shared differently with respect to their positions inside the tube wall. While the relaxation is more effective in the outter layer, the energy reduction in the inner In-rich layer is less pronounced. A similar trend is found in the GaAs layers, with the innermost layer of the tube less energetic than the one sandwiched between $In_{0.2}Ga_{0.8}As$ layers. The comparison between the red profile and the blue profile in Fig 9(b) shows how the lattice relaxation along the tube longitudinal axis is relevant for the overall energy minimization. The local decrease of energy in the In-rich layers is enough to compensate the increment that takes place in the GaAs layers. A more equitable energy partition is found in this configuration, with about 50% of the final elastic energy stored in both GaAs layers, that account nevertheless for 2/3 of the wall volume, and 50% in both $In_{0.2}Ga_{0.8}As$ inner and outter layers. The integrated energy along a radial atomic chain reduces from 568meV for the configuration with $a_l = a_{GaAs}$ to 460meV (for the flat films $E_{total}$ = 903meV).

Finally, the same effect of energy reduction due to longitudinal relaxation is observed for the dislocated layers. Considering the limit case of dislocated free films for this sample the vertically integrated elastic energy in one atomic chain of the flat $In_{0.33}Ga_{0.67}As$ film would be of 1290meV (8.1 meV/atom). Using the 0.7% biaxial in-plane relaxation which is suggested by the tube radius and in-plane/longitudinal lattice parameter such value would decrease to 580meV (3.6 meV/atom), which can be considered as a lower bound for the total dislocated bilayer elastic energy since locally the upper GaAs layer would have a tensile strain due to the partial relaxation of the In-rich film. Although it is hard to determine precisely the exact strain status of the flat layers Fig. 9(c) shows the energetic reduction due to the partially relaxed longitudinal lattice parameter in the dislocated tube. For the obtained strain profile and observed radius of the rolled-up structure the radially integrated energy is of 418meV while a much larger amount (596meV) would be expected for $a_l = a_{GaAs}$.

**V. Conclusion**

In this work we have shown how the strain distribution inside rolled-up tubes can be retrieved by x-ray micro-diffraction. Additionally, $\phi$-scans are able to reveal the overall mosaic spread of the tube layers. A simple x-ray kinematical model allows for a quantitative fitting of the diffraction curves. The unambiguous determination of values such as layer thickness, roughness, lattice parameter profiles and tube diameter can be performed due to the interconnected relations of continuum elasticity represented by equations 2-5. Radial lattice parameter profiles directly obtained from the x-ray diffraction allow for the determination of elastic deformation of the unit cell in rolled-up layers with an inherent precision of $10^{-3}$Å. From the fairly good model sensitivity to the longitudinal strain status shown here it was possible to infer that wafer-like interlayer bonding takes place in small radius tubes as shown for the bilayer and dislocated samples. In wide radius structures, such as the quad-layer tubes, the large distance from the rolled-up material to the un-etched front allows for a reduction of the elastic energy via longitudinal lattice relaxation. A precise knowledge of such elastic behavior on rolled-up layers with thicknesses of few tenths of nanometers is crucial to realize strain engineering for devices based on the rolling up principle [10, 12].

Authors acknowledge H. Djazouli, M. Weisser and E. Coric for technical support. Ch. D. likes to thank the ID01 staff for their hospitality during his stays at the beamline. This research is granted by the BMBF (project no. 03X5518).

## FIGURE CAPTIONS

Fig. 1 (color online) – (a) Schematic representation of the rolling-up of a bilayer semiconductor tube. (b) Sketch of the lattice parameter components and radius of a section from the inner turn of a bilayer tube [zoom of dashed rectangle area of (a)]. See text for discussion. (c) Elastic energy of an atomic chain in the radial direction – sketched as the dashed red rectangle in (b) – evaluated in the $R_i$ - $a_i$ parameter space for a 200Å $In_{0.2}Ga_{0.8}As$ / 200Å GaAs bilayer.

Fig. 2 (color online) – (a-c) Left panels: schematic layer layouts with nominal thickness of the samples used in this work; (a) bilayer, (b) quad-layer and (c) dislocated layer stack. X-ray reflectivity measurements (open dots) performed to extract the thickness of each flat layer stack are shown in the right panels of (a-c). The solid lines in all graphs are fits using Parratt32 [31]. SEM images of tubes obtained after rolling of the layers sketched in the left panels of a-c are shown in the insets.

Table I – Nominal and measured layer thickness for the flat and rolled-up layers. Values of average roughness $\sigma$ obtained from the fits of the reflectivity curves (Fig. 2) are also provided.

Fig. 3 (color online) – (a) Optical microscopy (left panel) and x-ray microdiffraction map (right panel) of the same region of the bilayer sample. (b) Tube diffraction geometry used for coplanar measurements detuned from the substrate lattice specular condition [25]. (c) Representation of the vicinity of the reciprocal space (004) diffraction position for flat layers (spots) and a cylindrical crystal rolled-up in a tube (rims). The red arrow illustrates schematically the reciprocal space path of a detuned radial scan.

Fig. 4 (color online) – Radial scans in the vicinity of the GaAs(004) reflection for the flat bilayer sample (blue dots) and the rolled-up layers (red dots). The blue line used for the flat layer data is a guide to the eyes (just connect dots). The partial strain relaxation that takes place after rolling can be observed as a shift in reciprocal space of the peaks in the red dots curve with respect to the GaAs bulk and pseudomorphically strained $In_{0.2}Al_{0.2}Ga_{0.6}As$ film positions. The inset shows schematically radially opposed regions of the tube that contribute to the x-ray diffraction signal [25].

Fig. 5 (color online) – (a) Representation of the x-ray diffracted beam footprint and the geometry of $\phi$-scans. (b) $\phi$-scans performed at the detuned GaAs (004) peak position for three tubes with different layer thickness. The red and green lines are lorentzian fits to scans in a 185Å (red open dots) and 300Å (green solid dots) GaAs layers in bilayer tubes (respectively). A similar scan with corresponding fit in the quad-layer tube (blue solid dots – blue line) is also shown.

Fig. 6 (color online) – (a) Radial scan measurement (dots) and simulation (line) for the bilayer tube. (b) Lattice parameter profiles in the inner, outter and middle layers of the 10-turn tube simulation used to fit the data of (a). (c) Fits using non-optimal parameters, used to extract effective intrinsic error bars of the modeling and infer the fit precision (see text for detailed explanation). (d) Lattice parameter profile for an In concentration of 0.21 in the In-rich layer. (e) Simulated lattice parameter profile for a value of $R_i$ 10% larger than the equilibrium value. The solid green and blue lines in (c) correspond to simulated diffraction profiles from (d) and (e).

Fig. 7 (color online) – (a) Radial scan near the GaAs(004) reciprocal space position for the quad-layer sample. The line is a guide to the eyes. (b) Radial scan (open dots) in a rolled-up tube from the quad-layer system. The dashed red curve represents a fit without longitudinal lattice parameter relaxation. For the blue solid line a relaxation of the lattice along the tube axis of 0.7% is assumed. (c) Lattice parameter profile for the quad-layer obtained from the fit of (b).

Fig. 8 (color online) – (a) Radial scans in the vicinity of the GaAs (004) reflection for the flat and rolled-up dislocated layers. The inset shows a grazing incidence radial scan close to the (400) in-plane reflection, allowing for the determination of the lattice parameter of the flat layers in the plane of the substrate. (b) Measured radial scan (dots) and fits to the dislocated tube diffraction. The orange solid line is obtained for the condition $a_l = a_{GaAs}$, while the blue line is the best fit assuming a 0.7% longitudinal lattice relaxation. The inset shows the lattice parameter profile that corresponds to the best fit.

Fig. 9 (color online) – (a) Local per-atom elastic energy profiles for the as-grown (black line) and rolled-up layers of the bilayer sample. The blue line represents the local elastic energy in the inner turn while the orange line represents the outter winding for a 10-turn tube. (b) Local elastic energy for the quad-layer flat films (black line), and rolled-up tube assuming no longitudinal lattice parameter relaxation (red lines) and the obtained lattice profile including the longitudinal relaxation given in Fig. 7(c) (blue lines). (c) Elastic energy evaluated locally for the dislocated bilayer tube, assuming $a_l=a_{GaAs}$ (red curve) and the 0.7% longitudinal relaxation (blue curve).

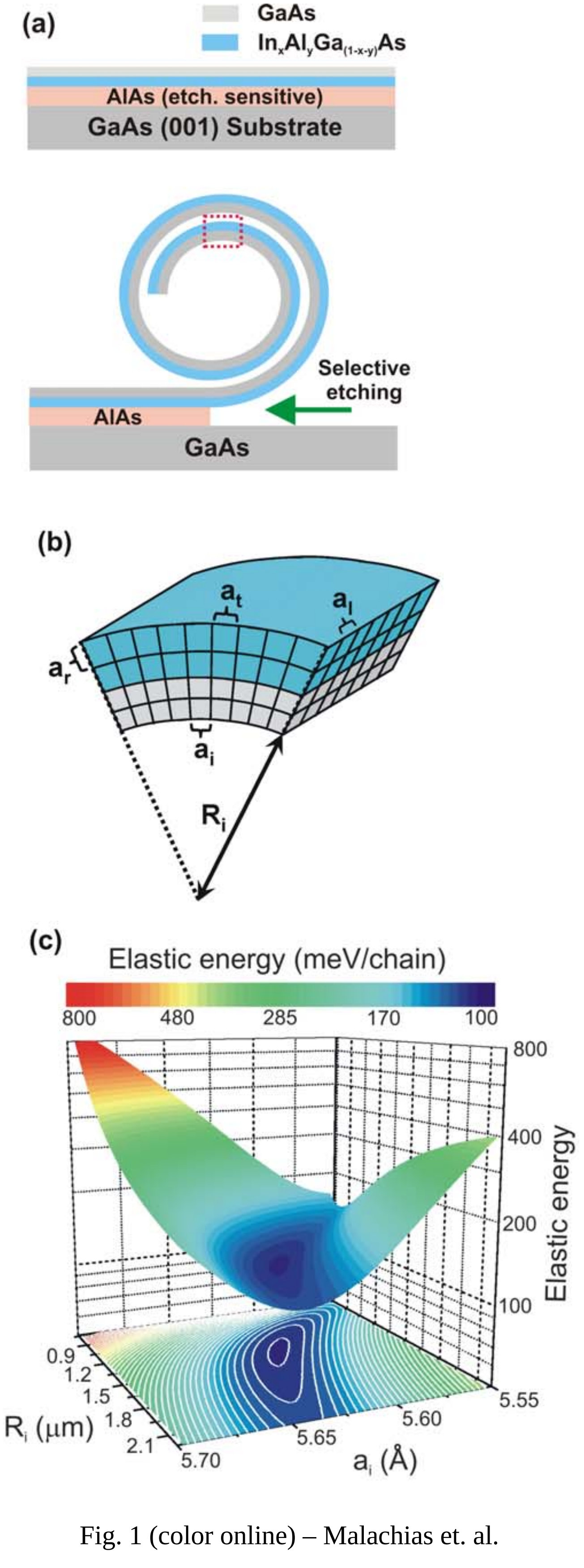

Fig. 1 (color online) – Malachias et. al.

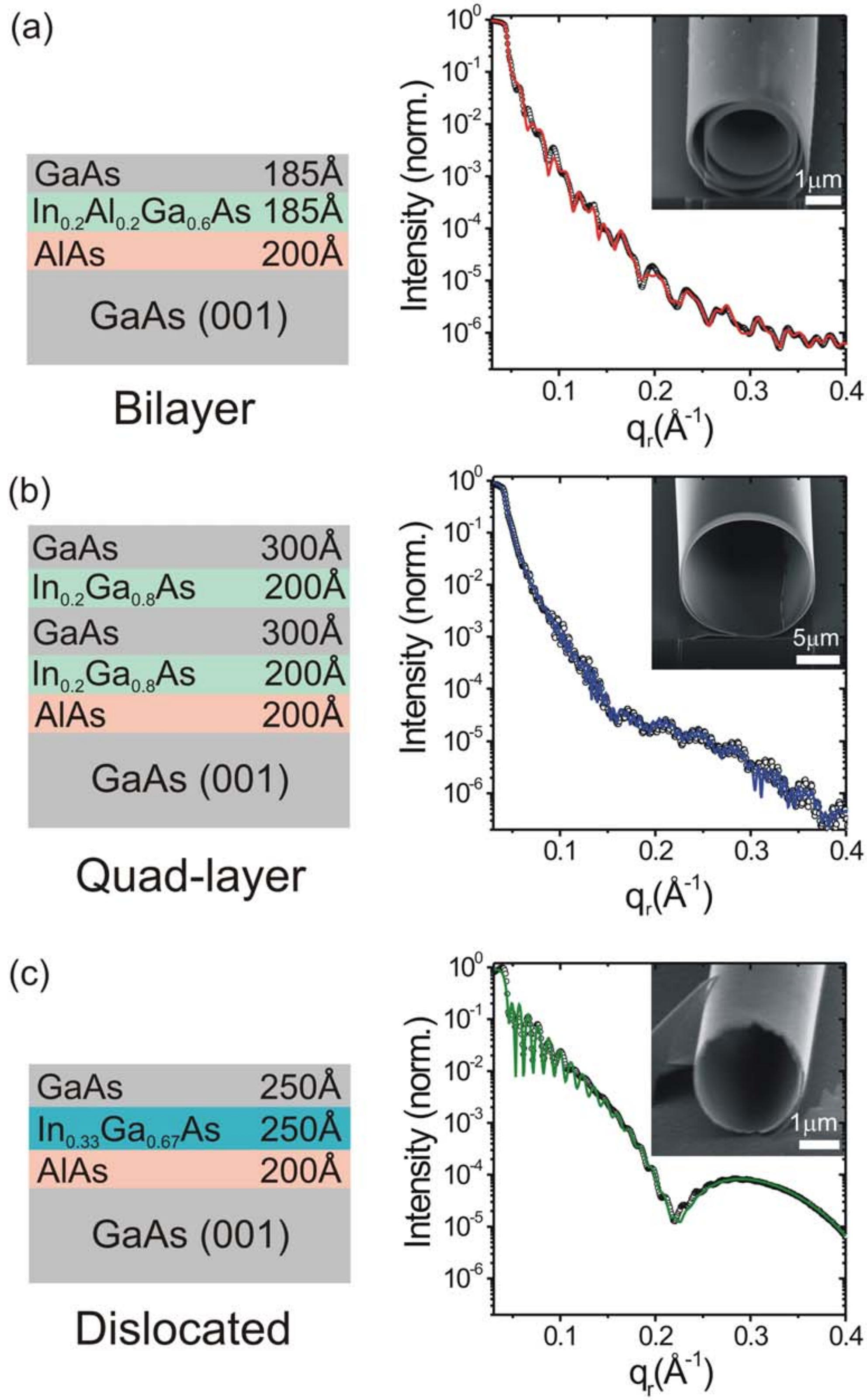

Fig. 2 (color online) – Malachias et. al.

| Sample | Layer | Nominal thickness | Reflectivity Flat layer thick. | Diffraction Rolled layer thick. |
|---|---|---|---|---|
| Bilayer | GaAs | 185Å | 198Å (σ = 2Å) | 190Å (σ = 20Å) |
| | $In_{0.2}Al_{0.2}Ga_{0.6}As$ | 185Å | 189Å (σ = 2Å) | 150Å (σ = 20Å) |
| Quad-layer | GaAs (top/inner) | 300Å | 285Å (σ = 10Å) | 280Å (σ = 30Å) |
| | $In_{0.2}Ga_{0.8}As$ (t/i) | 200Å | 196Å (σ = 9Å) | 190Å (σ = 30Å) |
| | GaAs (bottom/outer) | 300Å | 314Å (σ = 9Å) | 270Å (σ = 30Å) |
| | $In_{0.2}Ga_{0.8}As$ (b/o) | 200Å | 213Å (σ = 9Å) | 180Å (σ = 30Å) |
| Dislocated | GaAs | 250Å | 227Å (σ = 40Å) | 225Å (σ = 45Å) |
| | $In_{0.33}Ga_{0.67}As$ | 250Å | 237Å (σ = 48Å) | 230Å (σ = 45Å) |

Table I – Malachias et. al.

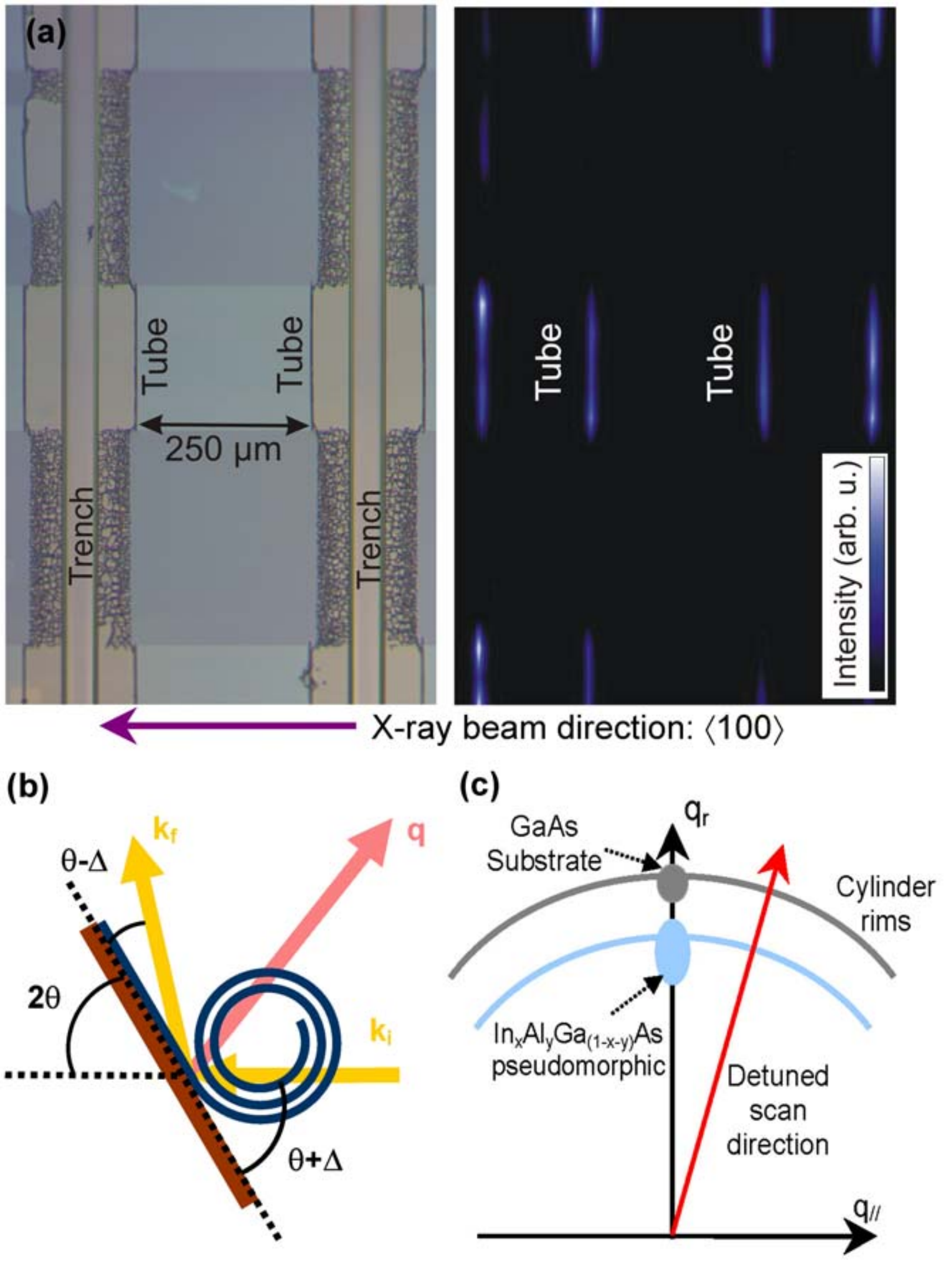

Fig. 3 (color online) – Malachias et. al.

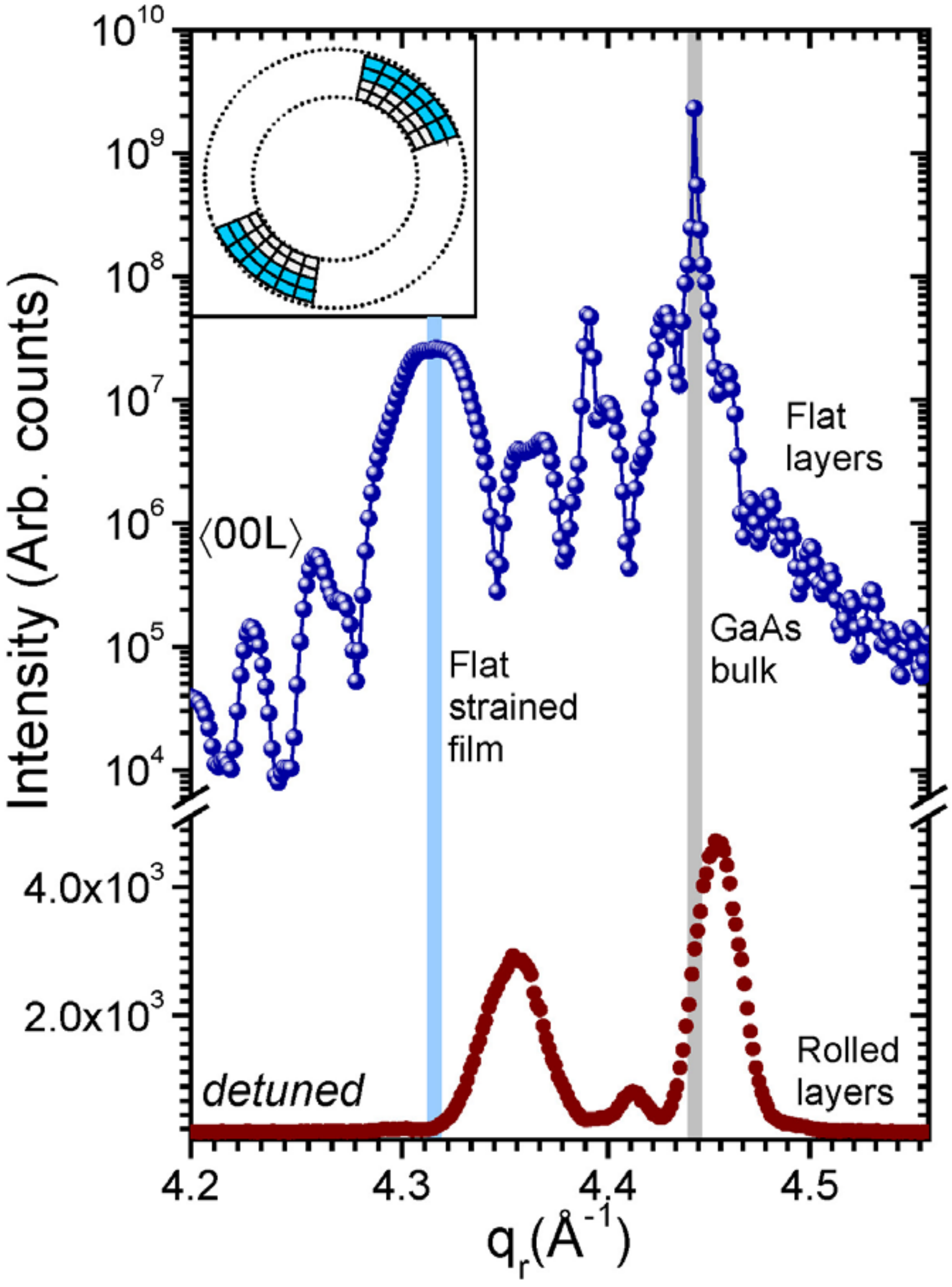

Fig. 4 (color online) – Malachias et. al.

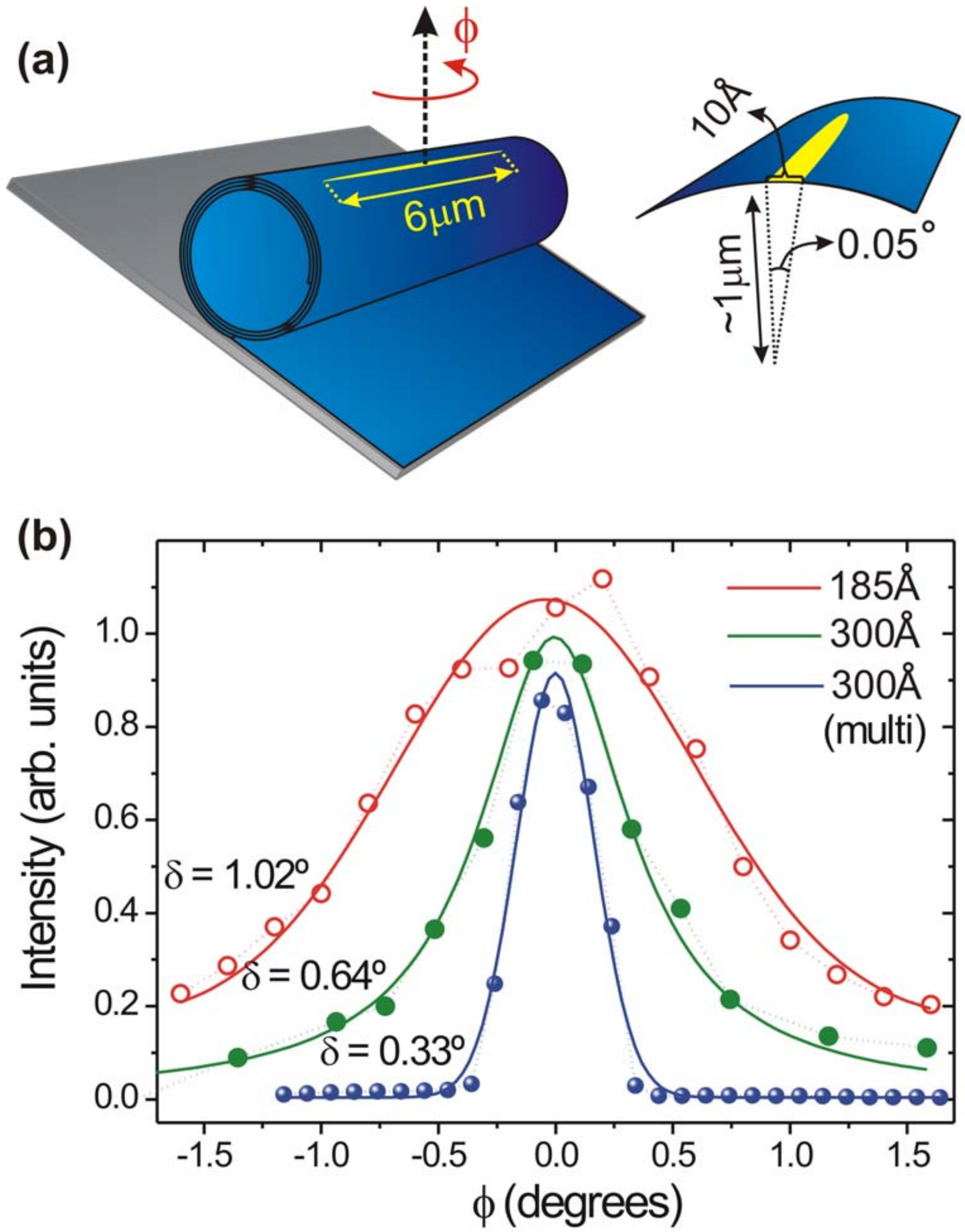

Fig. 5 (color online) – Malachias et. al.

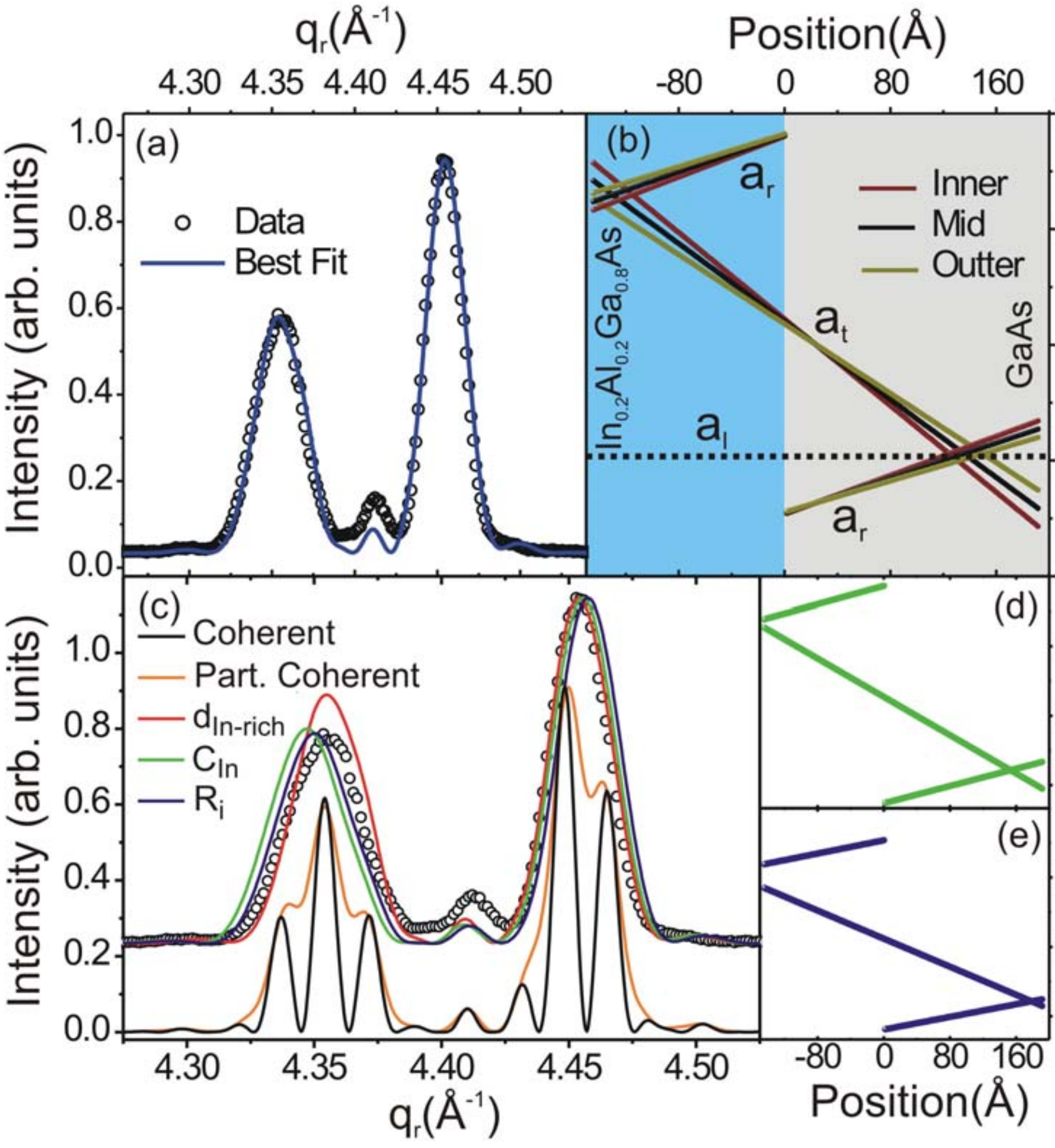

Fig. 6 (color online) – Malachias et. al.

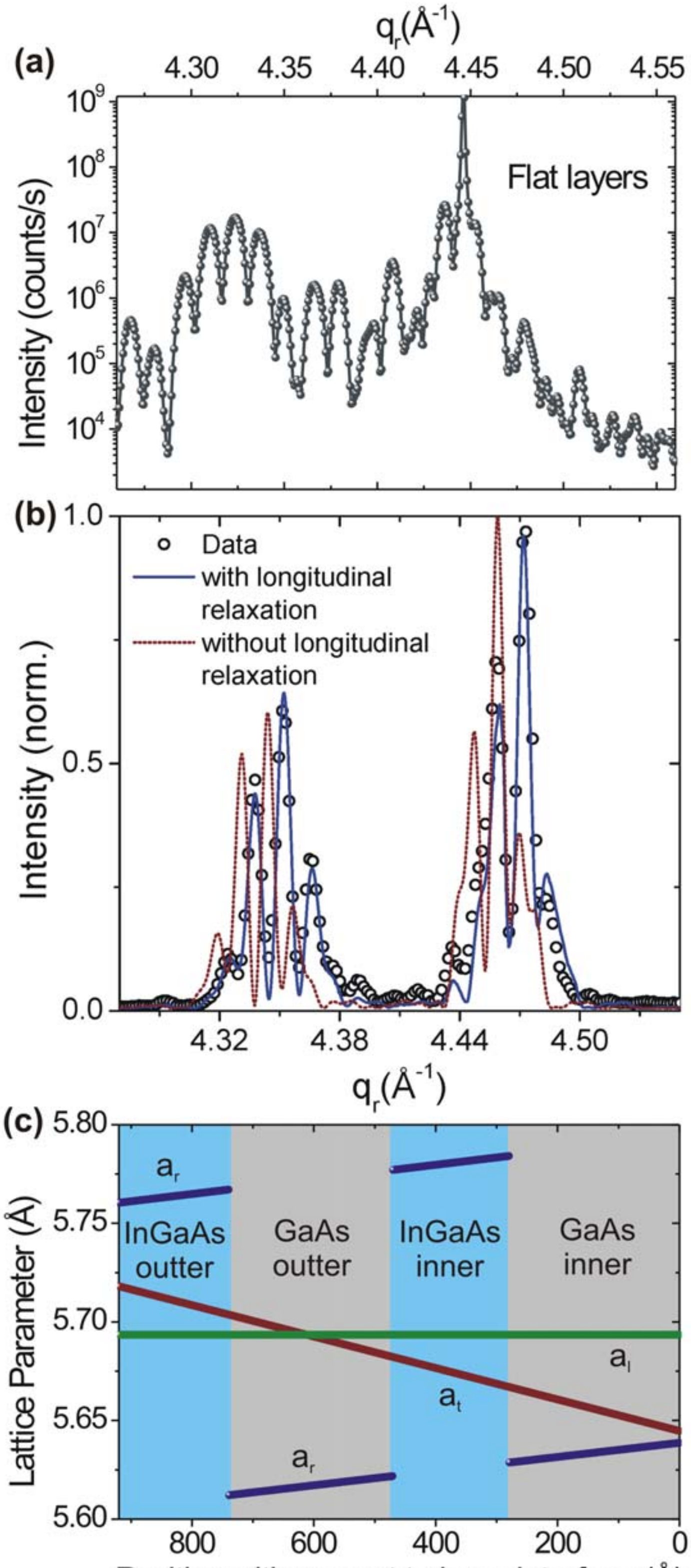

Fig. 7 (color online) – Malachias et. al.

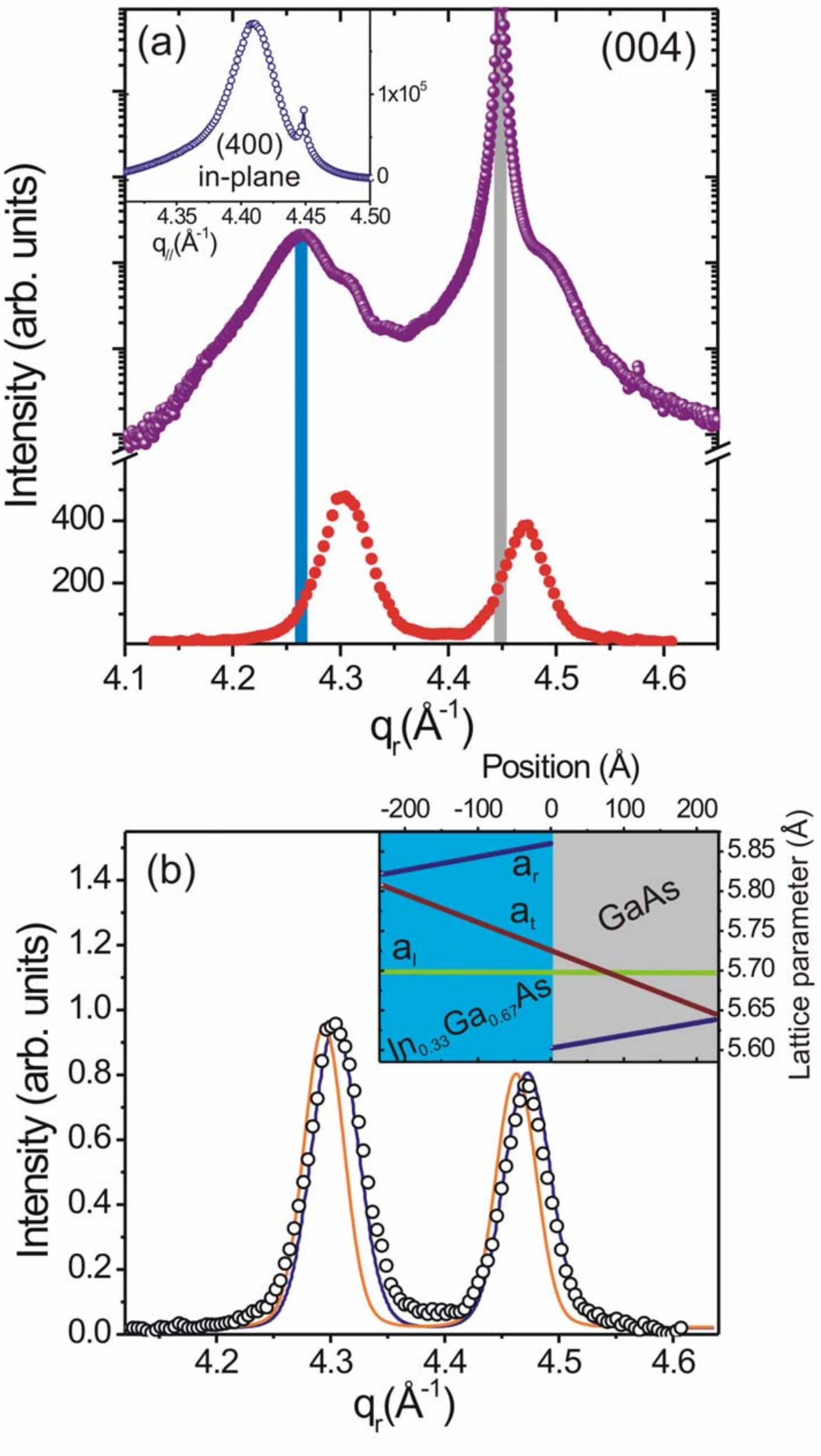

Fig. 8 (color online) – Malachias et. al.

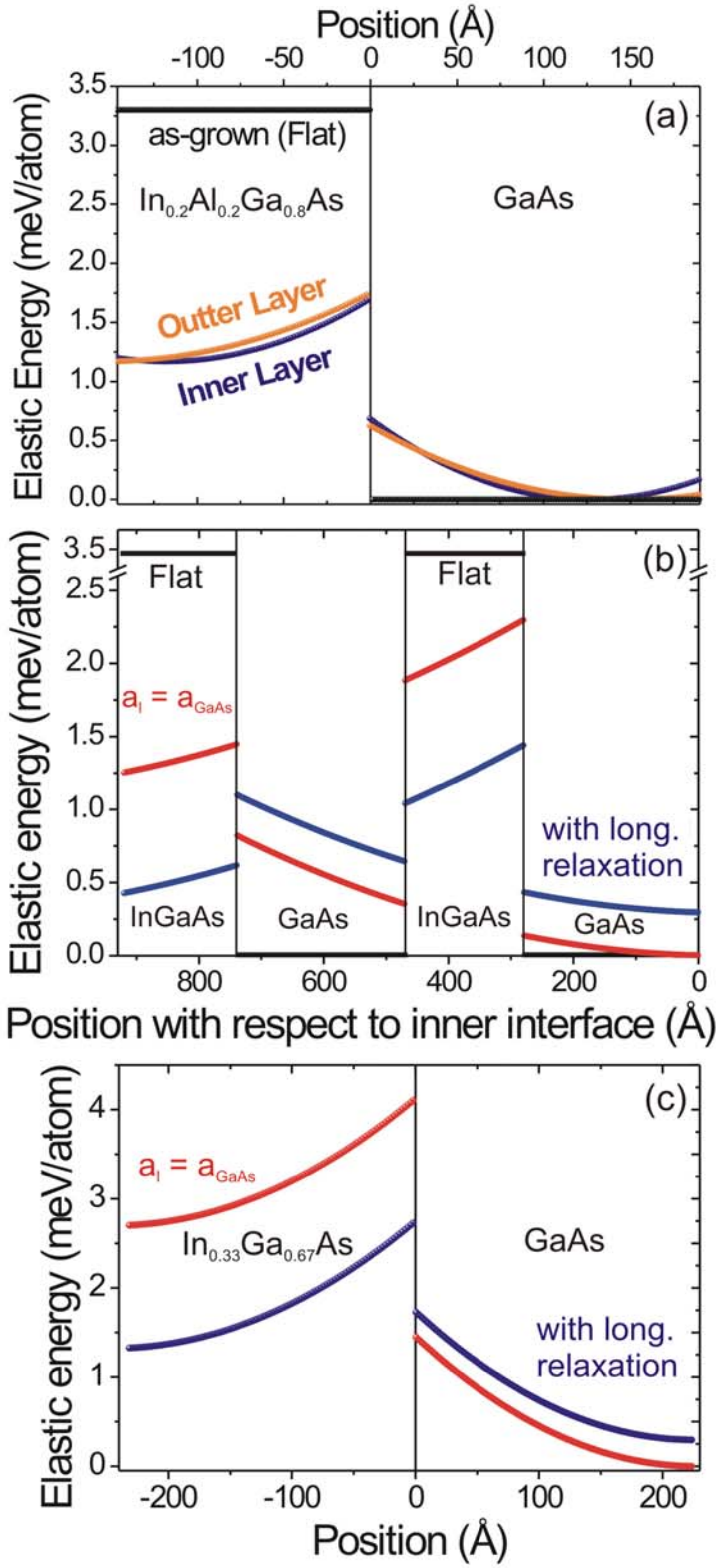

Fig. 9 (color online) – Malachias et. al.